\documentclass{iaus}
\usepackage{graphicx}

\newcommand{\aj}{AJ}
\newcommand{\apj}{ApJ}
\newcommand{\apjl}{ApJL}

\newcommand{\aap}{A\&Ap}

\newcommand{\mnras}{MNRAS}
\newcommand{\nat}{Nature}
\newcommand{\HIP}{{\it HIPPARCOS}}
\def\arcdeg{\hbox{$^\circ$}}
\newcommand{\msun}{$M_{\odot}$}

\title[IAUS 248.~~HST FGS Astrometry] 
{HST FGS astrometry - the value of fractional millisecond of arc precision}

\author[Benedict, McArthur, \& Bean]   
{G. Fritz Benedict$^1$, Barbara E. McArthur$^1$, \and Jacob L. Bean$^2$}

\affiliation{$^1$McDonald Observatory, University of Texas, Austin, TX USA 78712 \\ email: {\tt fritz or mca@astro.as.utexas.edu}\\  $^2$ Institute f\"{u}r Astrophysic, Georg-August-Universit\"{a}t, G\"{o}ttingen, Germany \\email: {\tt bean@astro.physik.uni-goettingen.de}}

\pubyear{2008}
\volume{248}  
\pagerange{119--126}
\jname{A Giant Step: from Milli- to Micro-arcsecond Astrometry
}
\editors{Jin Wenjing, ed.}
\begin{document}

\maketitle

\begin{abstract} 
In a few years astrometry with the venerable combination of Hubble Space Telescope and Fine Guidance Sensor will be replaced by SIM, GAIA, and long-baseline interferometry. Until then we remain a resource of choice for sub-millisecond of arc precision optical astrometry. As examples we discuss 1) the uses which can be made of our parallaxes of galactic Cepheids, and 2) the determination of perturbation orbital elements for several exoplanet host stars, yielding true companion masses. 
\end{abstract}

\firstsection 
\section{Introduction}

In the early 1970's, Bob O'Dell, then director of Yerkes Observatory, twisted the arm of Bill van Altena, encouraging Bill to propose to study the feasibility of doing astrometry with what was then called the Space Telescope (ST). Bill was joined by Larry Frederick and Otto Franz. Their efforts insured that astrometry with a Fine Guidance Sensor (FGS) would be an integral science component for ST. In 1977 Bill Jefferys formed a team to respond to the ST Announcement of Opportunity. The first author was a member of that team, which then included Raynor Duncombe, Paul Hemenway, and Pete Shelus. We were eventually chosen as a component (along with van Altena, Franz, and Fredrick) of the final STAT (Space Telescope Astrometry Team). Barbara McArthur joined us in 1985. 

ST was launched in 1990, becoming the Hubble Space Telescope (HST). Because of seismic disturbances caused by day/night cycle temperature effects on the original solar arrays and slow, but fortunately steady, changes in the FGS, we were unable to prove that precision astrometry could be carried out with the FGS until after the first refurbishment mission in late 1992. With new, re-designed solar arrays, HST and the two FGS units we have used over the years, originally predicted to produce single-measurement astrometry with 3 millisecond of arc (mas) precision,  have both routinely produced 1 mas or better per-observation precision astrometry. Calibration issues, many of which are non-trivial, are discussed in McArthur  \etal~(2002, 2006), Benedict  \etal~(1994, 2002a), and for binary star fringe morphology analysis, Franz  \etal~(1998). 

The many results from the original STAT include parallaxes of astrophysically interesting stars (Benedict \etal~ 2000, 2002a, 2002b, 2003, McArthur \etal~ 1999, 2001), a parallax for the Hyades (van Altena \etal~ 1997), a link between quasars and the HIPPARCOS reference frame (Hemenway \etal~ 1997), determination of low-mass binary star masses (Franz  \etal~~1998, Benedict  \etal~~2001), and searches for Jupiter-mass companions to Proxima Cen and Barnard's Star (Benedict \etal~ 1999). 

Since 1999, Barbara and I have carried on as General Observers, contributing to the study of the lower main sequence mass-luminosity relationship (Henry  \etal~1999), the intercomparison of dwarf novae (Harrison  \etal~2004), and parallaxes of cataclysmic variables (Beuermann  \etal~2003, 2004; Roelofs  \etal~2007) and
the Pleiades (Soderblom  \etal~2005). Jacob Bean, now a postdoc, joined us as a graduate student in 2002.

Two major themes of our investigations over the last few years have included the cosmic distance scale and extrasolar planetary systems. The remainder of this paper will outline recent results, specifically drawing from our work (Benedict  \etal~2007) on the galactic Cepheid Period-Luminosity Relationship (PLR), and the determination of extrasolar planetary masses (Benedict  \etal~2002c, McArthur  \etal~2004, Benedict  \etal~2006, Bean  \etal~2007).

\section{The Galactic PLR}
Our goal was to determine trigonometric parallaxes  for nearby fundamental mode Galactic Cepheid variable stars. Our  target selection consisted in choosing the nearest Cepheids (using \HIP~ parallaxes, Perryman \etal~1997), covering as wide a period range as possible. These stars were the brightest known Cepheids at their respective periods. Our new parallaxes provided distances and ultimately absolute magnitudes, $M$, in several bandpasses. Additionally, our investigation of the astrometric reference stars provided an independent estimation of the line of sight extinction to each of these stars, a contributor to the uncertainty in the absolute magnitudes of our prime targets.
These Cepheids, all with near solar metallicity, should be immune to variations in absolute magnitude due to metallicity variations, {\it e.g.} \cite{Gro04,Mac06}. Adding our previously determined absolute magnitude for $\delta$ Cep, \cite{Ben02b}, we established V, I, K, and W$_{VI}$ Period-Luminosity Relationships using ten Galactic Cepheids with average metallicity, $\langle$[Fe/H]$\rangle$=0.02, a calibration that can be directly applied to external galaxies whose Cepheids exhibit solar metallicity. Details are given in \cite{Ben07}.

\begin{figure}[h]
\begin{center}
\includegraphics[width=4in]{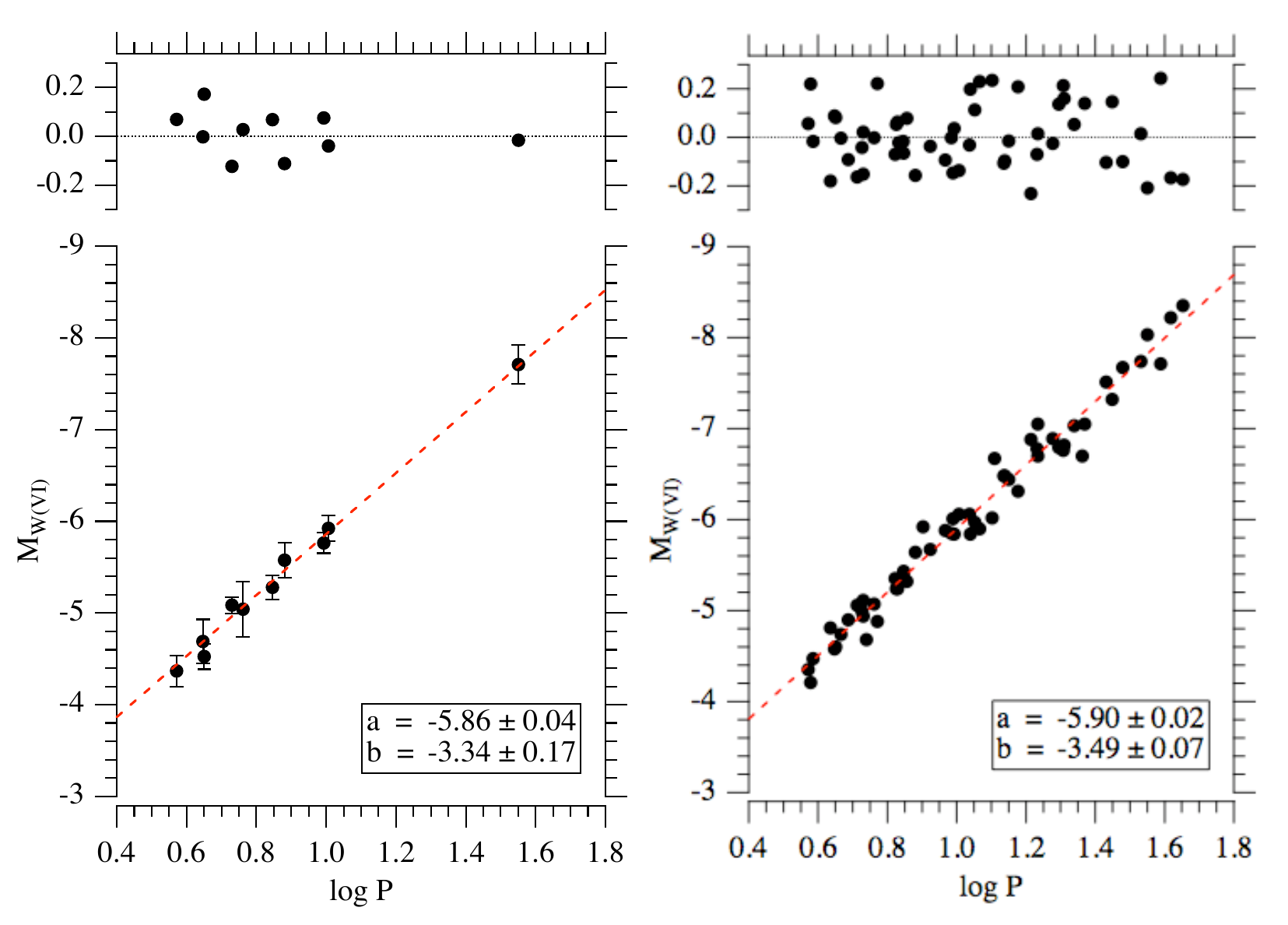} 
 \caption{The densification of the PLR. Our original Cepheid parallaxes (left) have calibrated techniques applied to many other, more distant Cepheids (right, after Fouqu\'{e}  \etal~~2007). Note the reduction in the error of the slope of the PLR.}
   \label{fig1}
\end{center}
\end{figure}

One of our interesting results is an independent determination of the distance modulus of the Large Magellanic Cloud (LMC), particularly given the recent exploration of possible socialogical influences on such determinations (Schaefer, 2007). OGLE (Udalski \etal~1999) has produced the largest amount of LMC Cepheid photometry. We worked with an apparent  W$_{VI}$ PLR for 581 Cepheids in the LMC. These data were carefully preened, selecting only Cepheids with normal light curves and amplitudes from  \cite{Nge05}. They  provide a zero-point of 12.65 $\pm$ 0.01. Direct comparison of that W$_{VI}$ zero-point with the zero-point of the left-hand M$_{W(VI)}$  PLR in Figure~\ref{fig1}  yields an LMC distance modulus 18.51 $\pm$0.03 with no metallicity corrections. 

 \cite{Mac06} demonstrate that a metallicity correction is necessary by comparing  metal-rich Cepheids with  metal-poor Cepheids in NGC 4258.  With a previously measured  [O/H] metallicity gradient from \cite{Zar94}, Macri  \etal~find a Cepheid metallicity correction in W$_{VI}$, $\gamma = -0.29 \pm 0.09_r \pm 0.05_s$ magnitude for 1 dex in metallicity, where r and s subscripts signify random and systematic. This value is similar
to an earlier W$_{VI}$ metallicity correction (Kennicutt et al. 1998) derived from  Cepheids in  M101 (-0.24$\pm$ 0.16).  Taking the weighted mean of the Kennicutt and Macri values and using the difference in metallicity of LMC and Galactic Cepheids (-0.36 dex from means of the data in
Groenewegen  \etal~2004 tables 3 and 4) we find a metallicity
correction of -0.10$\pm$ 0.03 magnitude with the Galactic Cepheids being
brighter. With this estimated metallicity
correction we obtain a corrected LMC modulus of 18.41 $\pm$ 0.05. One other recent determination is noteworthy for its lack of dependence on any metallicity corrections.  \cite{Fit03} derive 18.42$\pm$ 0.04, 
from eclipsing binaries, a modulus in close agreement with our new value. 

As an additional check on our adopted LMC modulus, m-M= 18.41 $\pm$ 0.05, a Macri  \etal~(2006) differential modulus (LMC-NGC 4258) leads to a modulus of 29.29 $\pm$ 0.08 for NGC 4258, a value in excellent agreement with the maser-based distance for NGC 4258, m-M = 29.29 $\pm$ 0.15 from \cite{Her99}.

One of the first uses of our new Cepheid parallaxes was the calibration of various techniques used to provide distances for many more long period (and all far more distant) Cepheids as described in \cite{Fou07}. These techniques include infrared surface brightness and interferometric Badde-Wesselink, which compare measured physical diameters with radial velocities. This calibration results in a densification and extension of the Galactic Cepheid PLR, and is illustrated in Figure~\ref{fig1}. The immediate value of the densification is a 
greater confidence in the absolute magnitudes of the longer-period, brighter Cepheids, the ones critical for extragalactic distance determination.

\section{Extrasolar Planet Masses}
 Currently, fewer than 10\% of the more than 200 candidate
exoplanets \\(http://exoplanet.eu) orbiting nearby stars have precisely determined
masses. Because the most successful technique for detecting candidate
exoplanets, the radial velocity method, suffers from a degeneracy between the mass and orbital inclination for most
of the known exoplanet candidates, only a minimum mass, $M sin i$, is known. In principle, radial velocities
alone can be used to determine the masses of exoplanets in multiplanet
systems, if two or more planets are experiencing significant mutual
gravitational interactions on short timescales. However, only one such
system has been investigated, results from different groups vary significantly,
and the effects of non-coplanarity remain to be considered.

Thus, establishing precise masses, rather than arguing
statistically (using $M sin i$, which everyone does, e.g. the otherwise excellent review of Udry \& Santos 2007), for the majority of exoplanet candidates requires additional observations. The primary techniques that
have been employed to break the mass-inclination degeneracy in
radial velocity data are astrometry (Benedict \etal~ 2002c, McArthur  \etal~2004, Benedict  \etal~2006, Bean \etal~ 2007) and
transit (e.g., Bakos \etal~ 2007) observations. Ideally, all candidates should have their masses determined. Obviously, not all systems are edge-on to us. And, unfortunately, the size of the perturbation due to the companion decreases linearly with distance, placing many of them beyond the reach of HST FGS astrometry.

\begin{table}[h]
\begin{center}
\caption{Masses from HST FGS Astrometry}
\begin{tabular}{|r|l|l|l|l|l|l|llllll} \hline
\multicolumn{1}{|r|}{Component}    &
 \multicolumn{1}{|l|}{M$_*$[\msun]}   &
 \multicolumn{1}{|l|}{[Fe/H]}   &
 \multicolumn{1}{|l|}{d (pc)}   &
 \multicolumn{1}{|l|}{ecc.}   &
  \multicolumn{1}{|l|}{$M_p$ ($M_{Jup}$)}   &
 \multicolumn{1}{|l|}{$\alpha$ (mas)}   &
 \multicolumn{1}{|l|}{inc. (\arcdeg)}   &
 \multicolumn{1}{|l|}{P (d)}   
\\ \hline
GJ 876 b$^1$&0.32&-0.12$^2$ &4.7&0.1&1.9$\pm$0.5&	0.25&84$\pm$6&61\\
55 Cnc d$^3$&1.21&0.32&12.5&0.33&4.9$\pm$1.1&1.9&53$\pm$7&4517\\
$\epsilon$ Eri b$^4$&0.83&-0.03&3.2&0.7&1.6$\pm$0.2&1.9&30$\pm$4	&2502\\
HD 33636	 B$^5$&1.02&-0.13&28.1&0.48&142$\pm$11&14.2&4.1$\pm$0.1&2117\\
\hline
\end{tabular}
\end{center}
\label{systab}
$^1$Benedict \etal~ 2002c, $^2$Bean \etal~ 2006a, 2006b, $^3$McArthur \etal 2004, $^4$Benedict \etal~ 2006, $^5$Bean \etal~2007\\
\end{table}

To partially rectify this lack of knowledge of the true masses, the HST allocation process awarded us over 100 orbits in Cycle 14 and 15 to be used to determine actual masses for HD 47536 b, HD 136118 b, HD 168443 c, HD 145675 b, HD 38529 c, and HD 33636 b. 
The object chosen for discussion in this contribution, HD 33636 (Bean \etal~2007) presents a cautionary tale. As we shall see, at least in some cases, $M sin i$ is not a very good estimator of the actual companion mass. 
\begin{figure}[b]
\begin{center}
\includegraphics[width=4in]{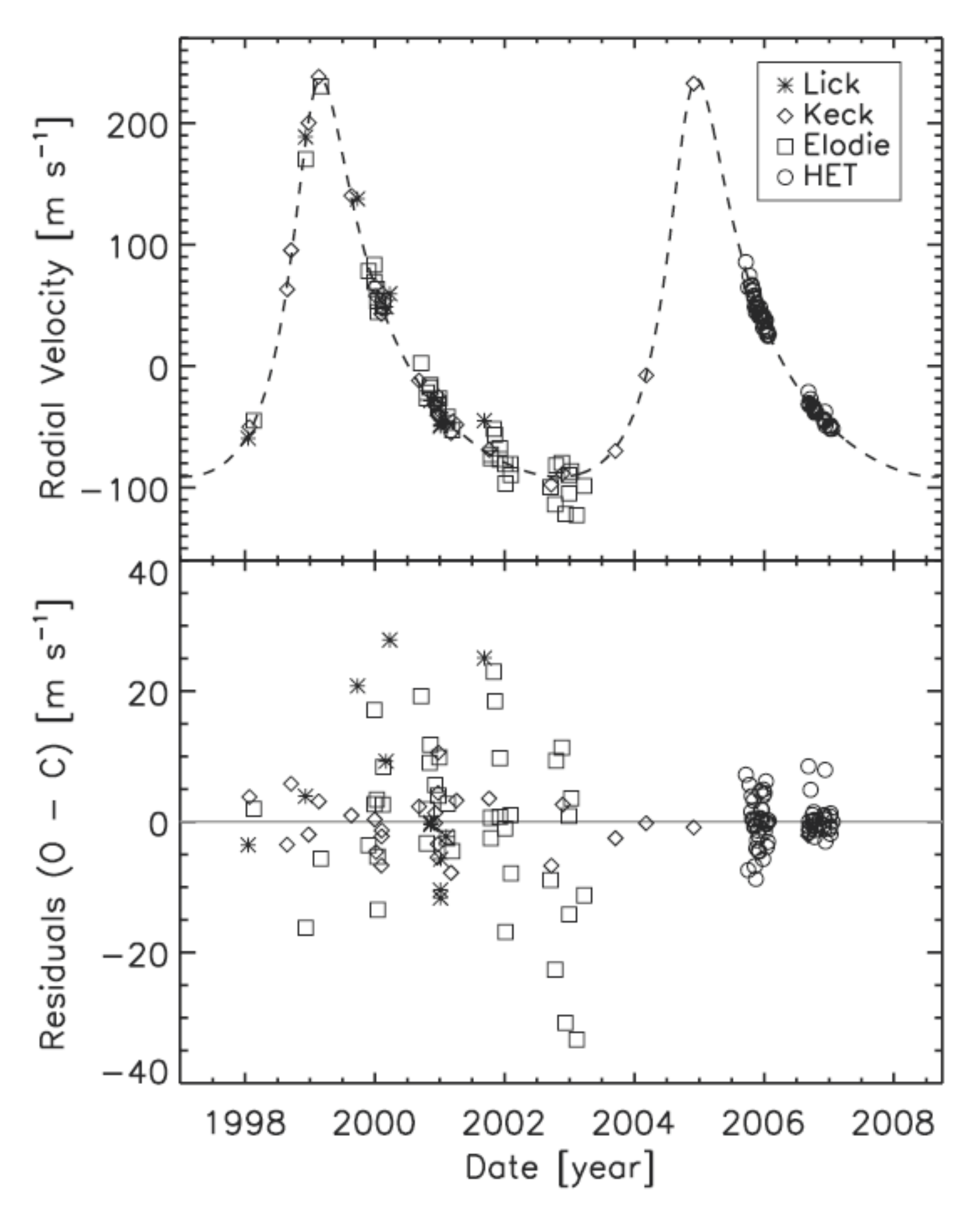} 
 \caption{HD 33636: Radial velocities and residuals to our final orbit from combined velocities and astrometry. The HET residuals are slightly better than the Keck residuals.}
   \label{fig2}
\end{center}
\end{figure}

We chose to work first on HD 33636 because we had finished  collecting all the HST data for this object. Radial velocity data (Figure 2) were obtained from past investigators (Lick and Keck data from Butler \etal~2006 and Marcy (2007, private communication), Elodie data from Perrier \etal~(2003)) and from a high-cadence campaign with the Hobby-Eberly Telescope High-Resolution Spectrograph (HRS; Tull 1998). The latter data were used to search for shorter-period companions, the unknown and unmodeled presence of which would have introduced noise into our velocity data. No such companions were found with detection limits detailed in Bean \etal~(2007). Limits depend on period and assumed eccentricity of the tertiary. For example these velocity data would have permitted detection of a tertiary with $M > 0.5M_{Jup}$ for $P < 300^d$ and eccentricity $<0.7$. 

Our techniques for combining radial velocity and HST FGS astrometry have been detailed in Benedict \etal~(2001, 2006) and Bean \etal~(2007). For HD 33636 we found an inclination, $i = 4.0\arcdeg \pm 0.1\arcdeg$ and, as shown in Figure 3, a perturbation semimajor axis, $\alpha = 14.2 \pm 0.2$ mas. Assuming a mass for the primary, HD 33636, $M = 1.02 \pm 0.03$ \msun ~(Takeda \etal~2007), we obtain a mass for HD 33636 b, $M = 142 \pm 11 M_{Jup} = 0.14 \pm 0.01$ \msun. HD 33636 b is actually HD 33636 B, an M6V star. Including it in any statistical studies depending on $M sin i$, such as metallicity, eccentricity, and semimajor axis, would corrupt those relationships.

\begin{figure}[b]
\begin{center}
\includegraphics[width=4in]{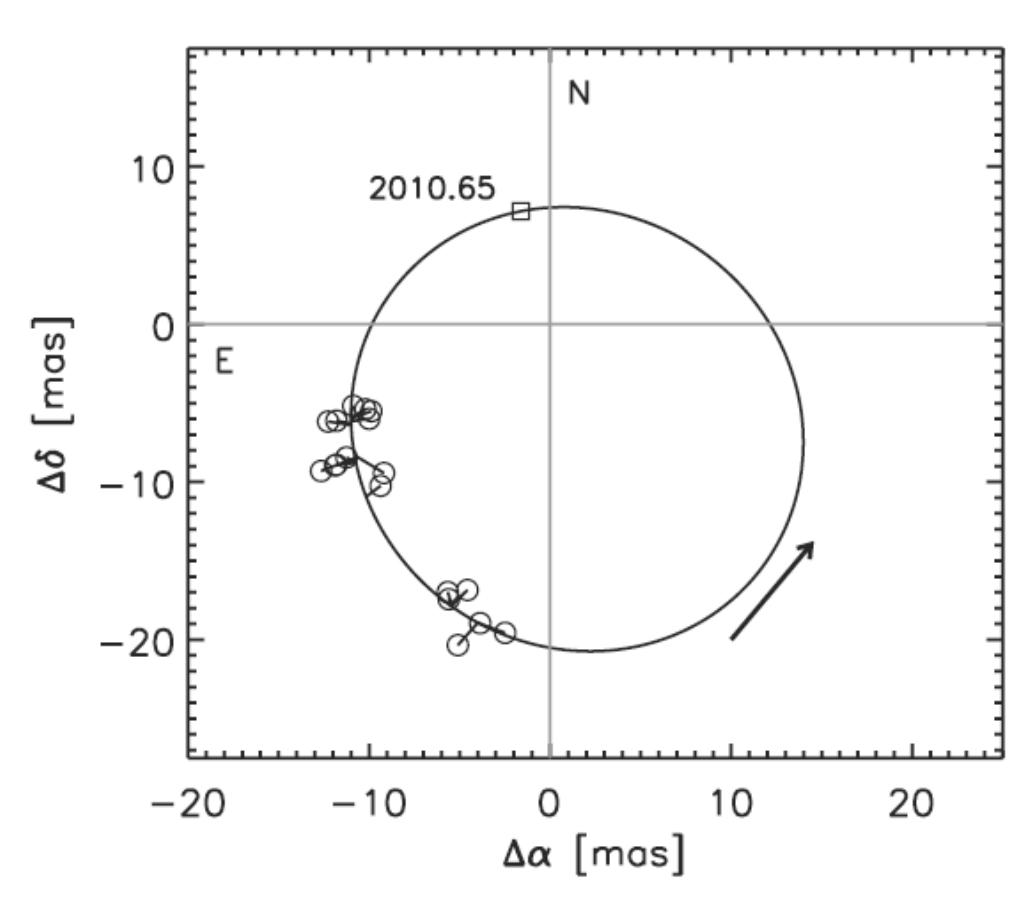} 
 \caption{The astrometric perturbation due to the {\em stellar} companion  to HD 33636. Based on radial velocities only this object had a minimum mass $M sin i= 9.4M_{Jup}$.
The inclination, $i = 4.1$\arcdeg, and size of the semi-major axis of the perturbation, $\alpha=14.2$ mas, identify the companion as an M dwarf star. The box marks the date of the next periastron passage.}
   \label{fig3}
\end{center}
\end{figure}

We have made some progress. Table 1 collects the mass estimates that have already resulted from our combination of HST astrometry and high-precision radial velocities.  Over the next year we will finish our analysis and determine actual masses for HD 47536 b, HD 136118 b, HD 168443 c, HD 145675 b, and HD 38529 c. But, before those results, we will have finished our analysis of the $\upsilon$ Andromedae system (McArthur \etal~2008). We have a preliminary indication that for the first time the degree of coplanarity of an extrasolar planetary system associated with a normal, main sequence star will be established. Beyond that, we are in the process of acquiring HST FGS data for similar coplanarity tests on the multiplanet systems associated with HD 128311, HD  202206, $\mu$ Ara, and $\gamma$ Cep, paving the way for future studies of hundreds of such systems with ground-based long-baseline interferometry and the space-based projects, SIM and GAIA.


\begin{discussion}

\discuss{audience}{}

\end{discussion}


\begin{thebibliography}{}




\bibitem[Bakos  \etal~(2007)]{Bak07} Bakos, G. A.  \etal~2007, \apj, 656, 552




\bibitem[Bean et al.(2006a)]{2006ApJ...653L..65B} Bean, J.~L., \etal~  2006a, \apjl, 653, L65 


\bibitem[Bean et al.(2006b)]{2006ApJ...652.1604B} Bean, J.~L. \etal~ 2006b, \apj, 652, 1604 

\bibitem[Bean et al.(2007)]{2007AJ....134..749B} Bean, J.~L. \etal~ 2007, \aj, 134, 749 

\bibitem[Benedict \etal(1998)]{Ben98} Benedict, G.~F. \etal~ 1998, \aj, 116, 429

\bibitem[Benedict \etal~(1999)]{Ben99} Benedict, G. F. \etal~ 1999, \aj, 118, 1086

\bibitem[Benedict \etal~(2000)]{Ben00} Benedict, G.~F. \etal~\ 2000, \aj, 119, 2382

\bibitem[Benedict \etal(2001)]{Ben01} Benedict, G.~F. \etal~ 2001, \aj, 121, 1607

\bibitem[Benedict  \etal~(2002a)]{Ben02a} Benedict, G. F. \etal~ 2002a, \aj, 123, 473

\bibitem[Benedict  \etal~(2002b)]{Ben02b} Benedict, G.~F., \etal.\ 2002b, \aj, 124, 1695 

\bibitem[Benedict et al.(2002c)]{Ben02c} Benedict, G.~F., et al.\ 2002c, \apjl, 581, L115 

 \bibitem[Benedict et al.(2006)]{2006AJ....132.2206B} Benedict, G.~F., et 
al.\ 2006, \aj, 132, 2206 

 \bibitem[Benedict \etal~ (2007)]{Ben07} Benedict, G.~F., \etal.\ 2007, \aj, 133, 1810 
 
\bibitem[Beuermann et al.(2003)]{2003A&A...412..821B} Beuermann, K. \etal~ 
2003, \aap, 412, 821
 
\bibitem[Beuermann et al.(2004)]{2004A&A...419..291B} Beuermann, K. \etal~
2004, \aap, 419, 291 




























\bibitem[Franz  \etal~(1998)]{Fra98} Franz, O.\ G. \etal~ 1998, \aj, 116, 1432

\bibitem[Fitzpatrick et al.(2003)]{Fit03} Fitzpatrick, E.~L. \etal~\ 2003, \apj, 587, 685 

\bibitem[Fouqu\'{e}  \etal~(2007)]{Fou07} Fouqu\'{e}, P.,  \etal~
2007, A\&A, 476, 73 











\bibitem[Groenewegen  \etal~(2004)]{Gro04} Groenewegen, M.~A.~T. \etal~ 2004, \aap, 420, 655 
 

\bibitem[Harrison \etal~ (1999)]{Har99} Harrison, T. E. \etal~1999, \apjl, 515, L93

\bibitem[Harrison et al.(2004)]{2004AJ....127..460H} Harrison, T.~E. \etal~ 2004, \aj, 127, 460 



\bibitem[Herrnstein \etal(1999)]{Her99} Herrnstein, J.~R., \etal\ 1999, \nat, 400, 539 



\bibitem[Kennicutt \etal~(1998)]{Ken98} Kennicutt, R.~C.~\etal\ 1998, \apj, 498, 181

 
 

 


 



\bibitem[Macri  \etal~(2006)]{Mac06} Macri, L.~M. \etal~ 2006, ApJ, 652, 1133





\bibitem[McArthur  \etal~(1997)]{McA97} McArthur, B. \etal~ 1997. in Proc. 1997 HST Calibration Workshop, ed. S. Casertano, R. Jedrzejewski, T. Keyes, and M. Stevens, STScI Publication, Baltimore, MD

\bibitem[McArthur  \etal~(1999)]{mca99} McArthur, B.~E. \etal~1999, \apjl, 520, L59 

\bibitem[McArthur  \etal~(2001)]{mca01} McArthur, B.~E. \etal~2001, \apj, 560, 907 

\bibitem[McArthur  \etal~(2002)]{McA02} McArthur, B. \etal~2002, The 2002 HST Calibration Workshop, Edited by Santiago Arribas, Anton Koekemoer, and Brad Whitmore.~Baltimore, MD: Space Telescope Science Institute, 2002, p.373 

\bibitem[McArthur et al.(2004)]{2004ApJ...614L..81M} McArthur, B.~E., et 
al.\ 2004, \apjl, 614, L81 

\bibitem[McArthur et al.(2006)]{2006hstc.conf..396M} McArthur, B.~E. \etal~ 2006, The 2005 HST Calibration Workshop: Hubble After the Transition to Two-Gyro Mode, 396 

\bibitem[McArthur et al.(2008)]{McA2008} McArthur, B.~E., \etal~ 2008, in preparation










\bibitem[Ngeow  \etal~(2005)]{Nge05} Ngeow, C.-C. \etal~ 2005, \mnras, 363, 831 





\bibitem[Perryman \etal(1997)]{Per97} Perryman, M.~A.~C. \etal~ 1997, \aap, 323, L49 




\bibitem[Roelofs et al.(2007)]{2007ApJ...666.1174R} Roelofs, G.~H.~A. \etal~ 2007, \apj, 666, 1174 









\bibitem[Schaefer(2007)]{2007arXiv0709.4531S} Schaefer, B.~E.\ 2007, ArXiv 
e-prints, 709, arXiv:0709.4531 



\bibitem[Soderblom  \etal~(2005)]{Sod05} Soderblom, D.~R. \etal~ 2005, \aj, 129, 1616 





\bibitem[Takeda \etal~(2007)]{Tak07} Takeda, G.,\etal~2007, ApJS, 168, 297

\bibitem[Tull(1998)]{1998SPIE.3355..387T} Tull, R.~G.\ 1998, Proc. SPIE, 
3355, 387 

\bibitem[Udalski  \etal~(1999)]{Uda99} Udalski, A. \etal~1999, Acta Astronomica, 49, 201 


\bibitem[Udry \& Santos~(2007)]{Udr07} Udry, S. \& Santos, N. C., 2007, Ann. Rev. Astron. \& Astrophys., 45, 379.



\bibitem[van Altena et al.(1997)]{1997ApJ...486L.123V} van Altena, W.~F., 
et al.\ 1997, \apjl, 486, L123 

\bibitem[van Leeuwen et al.(2007)]{2007MNRAS.379..723V} van Leeuwen, F. \etal~2007, \mnras, 379, 723 




\bibitem[Zaritsky \etal(1994)]{Zar94} Zaritsky, D. \etal~ 1994, \apj, 420, 87 
\end{thebibliography}
\end{document}